\documentclass[aps,prb,amsmath,amssymb,reprint,superscriptaddress,preprintnumbers,showpacs,intlimits]{revtex4-1}
\usepackage{bm,latexsym,mathrsfs,enumerate,color}
\usepackage[mathcal]{euscript}
\usepackage[breaklinks=true,unicode=true,urlcolor = blue,colorlinks = true,citecolor = blue,linkcolor = blue]{hyperref}
\usepackage{graphicx}

\renewcommand{\vec}[1]{\bm{#1}}
\begin{document}


\title{Periodic magnetic structures generated by spin-polarized currents in nanostripes}

\author{Oleksii M. Volkov}
     \email{alexey@volkov.ca}
\affiliation{Taras Shevchenko National University of Kiev, 01601 Kiev, Ukraine}

\author{Volodymyr P. Kravchuk}
 \email{vkravchuk@bitp.kiev.ua}
 \affiliation{Bogolyubov Institute for Theoretical Physics, 03143 Kiev, Ukraine}

\author{Denis D. Sheka}
 \email{sheka@univ.net.ua}
\affiliation{Taras Shevchenko National University of Kiev, 01601 Kiev, Ukraine}

\author{Franz G. Mertens}
\email{franzgmertens@gmail.com}
\affiliation{Physics Institute, University of Bayreuth, 95440 Bayreuth, Germany}

\author{Yuri~Gaididei}
 \email{ybg@bitp.kiev.ua}
 \affiliation{Bogolyubov Institute for Theoretical Physics, 03143 Kiev, Ukraine}

\date{\today}

%
%

\begin{abstract}
The influence of a spin-polarized current on long ferromagnetic nanostripes is studied numerically. The current flows perpendicularly to the stripe. The study is based on the Landau-Lifshitz phenomenological equation with the Slonczewski-Berger spin-torque term. The magnetization behavior is analyzed for all range of the applied currents, up to the saturation. It is shown that the saturation current is a nonmonotonic function of the stripe width. For a stripe width increasing it approaches the saturation value for an infinite film. A number of stable periodic magnetization structures are observed below the saturation. Type of the periodical structure depends on the stripe width. Besides the one-dimensional domain structure, typical for narrow wires, and the two-dimensional vortex-antivortex lattice, typical for wide films, a number of intermediate structures are observed, e.g. cross-tie and diamond state. For narrow stripes an analytical analysis is provided.
\end{abstract}




\maketitle

A magnetic nanostripe is a convenient system for studying the dynamics of magnetization structures driven by a spin-polarized current. Typically the current is passed along the stripe, which causes a movement of the domain wall. This phenomenon is widely studied both theoretically and experimentally, see e.g. reviews~\onlinecite{Lindner10,Tatara08,Marrows05,Klaui08}. Nevertheless the influence of a perpendicular current on the stripe magnetization dynamics is also of high interest for spintronic applications. Recently it was predicted theoretically~\cite{Khvalkovskiy09} and later confirmed experimentally~\cite{Boone10a,Uhlir2010,Chanthbouala11,Metaxas2013} that the perpendicular current can excite the domain wall motion with much higher velocity comparing to the in-plane current.

\begin{figure}
\includegraphics[width=0.8\columnwidth]{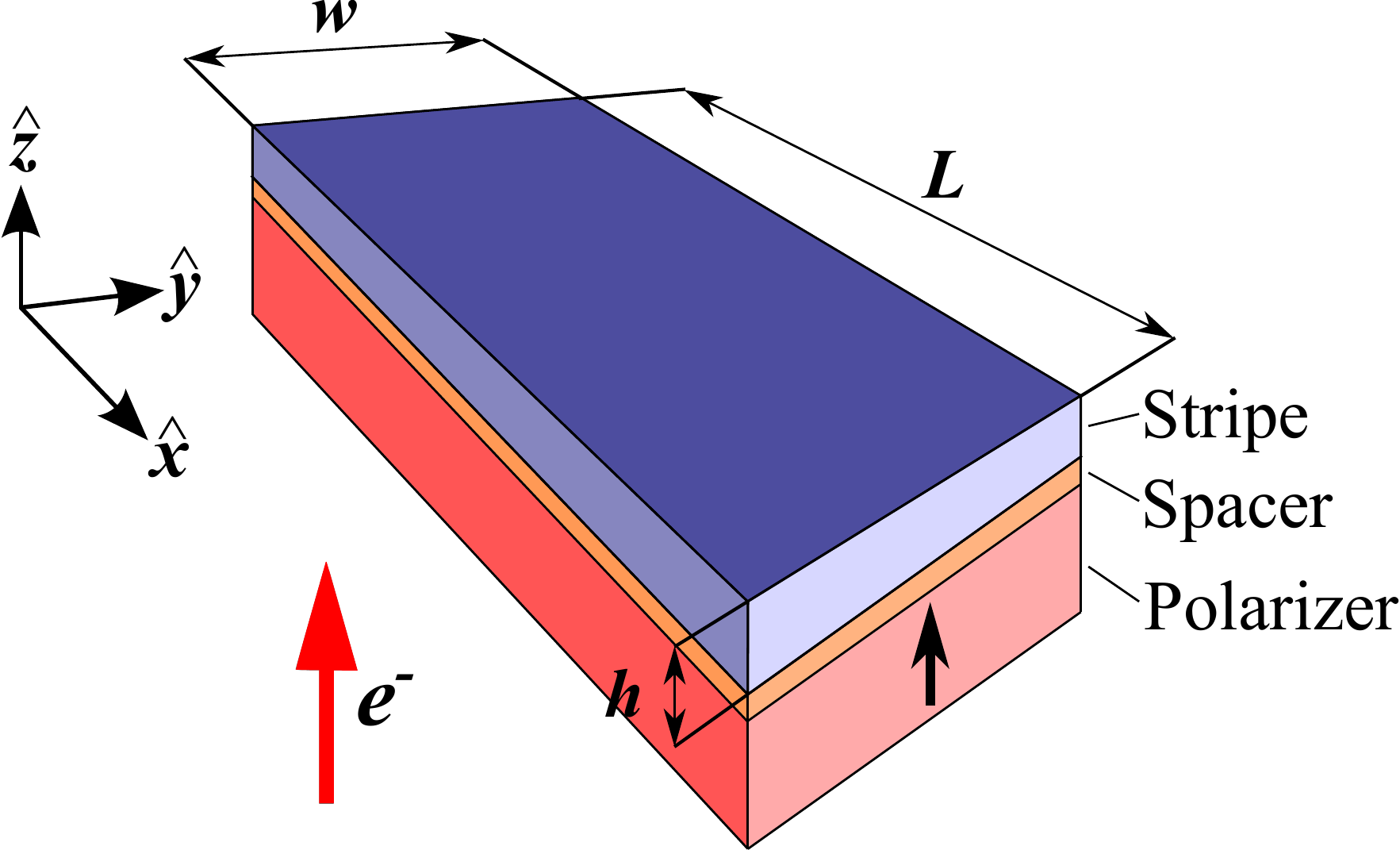}
\caption{(Color online) The three-layer stripe-shaped spin valve. The spin polarized current flows perpendicularly to the studied stripe opposite to $\hat{z}$-direction, thereby the conduction electrons flow in the opposite side, as  shown by the red (large) arrow. The black (small) arrow indicates the direction of the polarizer magnetization. }\label{fig:coords}
\end{figure}

Recently we have studied the action of the strong perpendicular spin-polarized current on a nanomagnet for two limit cases, namely a planar two-dimensional film\cite{Volkov11,Gaididei12a} and a narrow one-dimensional wire\cite{Kravchuk2013}. In both cases a stable periodical structure induced by the spin-current is found in the pre-saturated regime: a square vortex-antivortex lattice is formed in a film and a one-dimensional domain structure is formed in a wire. The aim of this paper is to make a link between these limit cases. For this purpose we consider thin and long stripe shaped samples of different widths. By varying the stripe width we study the current induced magnetization behavior in wide range, starting from quasi-one dimensional narrow strips ($w/h\lesssim1$) and up to quasi two-dimensional wide strips ($w/h\gg1$), where $w$ and $h$ denote respectively the stripe width and thickness. The following analysis is made under the assumption that the stripe is sufficiently long, so that $L\gg w$ and $L\gg h$ with $L$ being the stripe length. We also assume that the stripe is thin enough to ensure uniformity of the magnetization along the thickness. Details of the problem geometry are shown in Fig.~\ref{fig:coords}.

Our study is based on the Landau-Lifshitz-Slonczewski phenomenological equation\cite{Slonczewski96,Berger96,Slonczewski02}:
\begin{equation} \label{eq:LLS}
\dot{\vec{m}} = \vec m\times{\delta\mathcal{E}}/{\delta\vec{m}} - j \varepsilon \vec m\times[\vec m\times\hat{\vec{z}}],
\end{equation}
where $\vec{m}=\vec{M}/M_s=(m_x, m_y, m_z)$ is the normalized magnetization, $M_s$ is the saturation magnetization. The overdot indicates a derivative with respect to the rescaled time in units $(4 \pi \gamma M_s)^{-1}$, $\gamma$ is a gyromagnetic ratio and $\mathcal{E}=E/(4\pi M^2_s)$ is the normalized magnetic energy. We consider here a soft ferromagnet, therefore only exchange and magnetostatic contributions to the total energy are taken into account. The normalized spin-current density $j=J/J_0$, where $J_0=4 \pi M^2_s|e|h/\hbar$, with $e$ being the electron charge, $\hbar$ is the Planck constant. The physical meaning of the quantity $J_0$ was clarified in the Ref.~\onlinecite{Volkov2013}, where it was shown that for currents $J\ge J_0$ a ``rigid'' saturation appears: the saturated state remains stable in a transverse magnetic field regardless of the field amplitude and direction. The spin-transfer torque efficiency function $\varepsilon$ has the form $\varepsilon= P \Lambda^2/\left[ (\Lambda^2+1)+(\Lambda^2-1) (\vec{m}\cdot\hat{\vec z})\right]$, where $P$ is the degree of spin polarization and the parameter $\Lambda$ describes the resistance mismatch between the spacer and the ferromagnet stripe\cite{Slonczewski02,Sluka11}. The damping was omitted in Eq.~\eqref{eq:LLS}, because, as it was shown earlier\cite{Gaididei12a,Kravchuk2013}, the transverse spin-polarized current plays the role of an effective damping, which is usually greater than the natural one. It should also be noted that the Eq.~\eqref{eq:LLS} is written for the case when the Polarizer is magnetized along the $z$-axis, see Fig.~\ref{fig:coords}.

Here we report on the results of a numerical study based on the micromagnetic simulations.\footnote{We use the OOMMF code, version 1.2a5 [http://math.nist.gov/oommf/] for material parameters of Permalloy ($\mathrm{Ni}_{81}\mathrm{Fe}_{19}$): saturation magnetization $M_s=8.6 \times 10^5$ A/m, exchange constant $A=13 \times 10^{-12}$ J/m, and  anisotropy is neglected. Size of the mesh cell is $3 \times X \times h$ nm, where $X$ takes values in interval from 2 to 3 nm, depending on the stripe width. The width is changes with steps $\Delta w=0.5$~nm for narrow stripes (0.5$\le w\le5$~nm) and $\Delta w=1$~nm for other samples (5$< w\le100$~nm). The current parameters are the following: polarization degree $P = 0.4$, and $\Lambda = 2$.} The length of all studied stripes is the same $L=1\,\mathrm{\mu m}$. To ensure the magnetization uniformity along the $z$-axis we consider only sufficiently thin stripes with a thicknesses not exceeding several characteristic magnetic length, namely $h$=5, 10, and 15~nm. The width is varied in a wide range 0.5$\le w\le100$~nm.  As an initial state for each simulation we choose a uniform in-plane magnetization along the stripe (along the $x$-axis), which is very close to the ground state of a long stripe. To consider all possible current values we adiabatically increase the current density until the stipe reaches the saturated state, when all magnetic moments are aligned along the $z$-axis.\footnote{Density of the applied current is changed accordingly to the law: $J=t~\Delta J/\Delta t$, where $\Delta J=10^{11}~A/m^2$ and $\Delta t=1~ns$. As a criterion of the saturation we use the relation $M_z/M_s>0.9999$, where $M_z$ is the total magnetization along the $z$-axis.}

\begin{figure*}
\begin{center}
\includegraphics[width=0.9\textwidth]{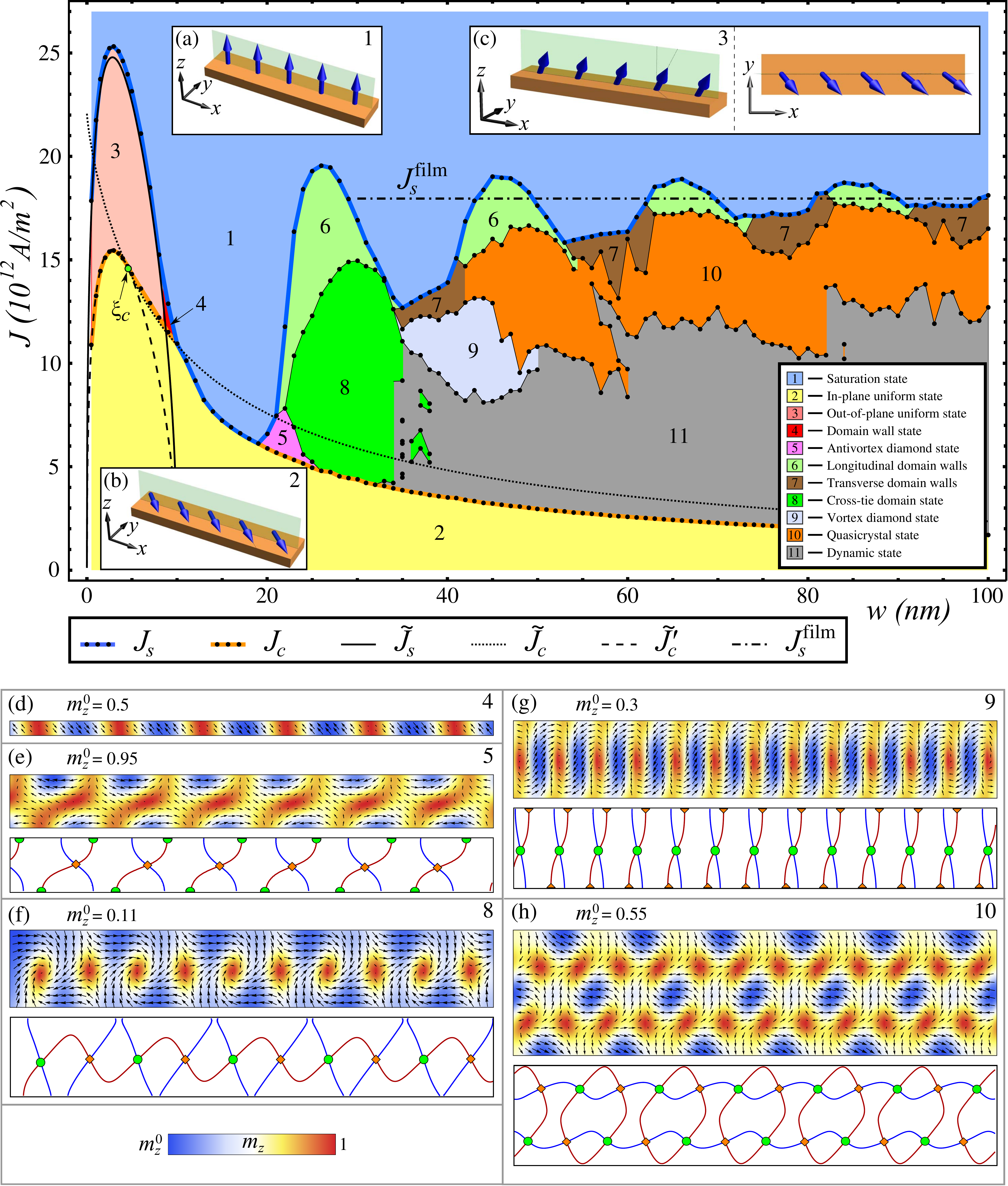}
\end{center}
\caption{(Color online) The phase diagram of the magnetisation behavior of ferromagnetic stripes of different widths $w$ under the action of a transverse spin current $J$. Length $L=1$~$\mu$m and thickness $h=10$~nm of the stripes are fixed. Black bold dots indicates the transition from one state to another, and each state is numerated, as shown on the inset. The thin horizontal dash-dot line indicates the saturation current $J_s^\mathrm{film}$ for an infinite film of thickness $h$. Black dotted and solid lines indicate analytically obtained currents $\tilde J_c$ and $\tilde J_s$, respectively, see Eqs.~\eqref{eq:J_c} and \eqref{eq:Js-tild}. The current $\tilde J_c'$ is shown by the dashed line. The insets (a), (b) and (c) show possible uniform states. Examples of possible periodic structures are shown below, the area of view is restricted to the 100~nm long central part of a stripe. Numbers of the insets corresponds to the numbers in the legend: (d) is the periodical domain wall structure ($w=9~nm$, $J=12 \times 10^{12}~A/m^2$), (e) is the antivortex diamond state ($w=22~nm$, $J=7.5 \times 10^{12}~A/m^2$), (f) is the cross-tie domain wall ($w=28~nm$, $J=7.5 \times 10^{12}~A/m^2$), (g) is the vortex diamond state ($w=40~nm$, $J=11.25 \times 10^{12}~A/m^2$), (h) is the quasicrystal structure ($w=52~nm$, $J=14 \times 10^{12}~A/m^2$). To determine positions of the vortices and antivortices we used the method\cite{Hertel07} of intersection of isolines $m_x=0$ (orange line) and $m_y=0$ (blue line). Circles and diamonds shows positions of vortices and antivortices respectively. Numbers on the insets correspond to the numbers of the regions on the diagram.}\label{fig:diagram}
\end{figure*}

All possible types of the magnetization behavior induced by spin-currents in different stripes can be summarized in form of the phase diagram presented in Fig.~\ref{fig:diagram}. First of all, one should distinguish two critical currents, namely the saturation current $J_s(w)$ which is a minimal current that takes the stripe to the saturated state, see the inset (a) in Fig.~\ref{fig:diagram}; and the current $J_c(w)$ which is the highest current at which the uniform in-plane state remains stable.\footnote{Since the diagram of states (see Fig.~\ref{fig:diagram}) is built for a certain thickness, we omit the dependence on $h$ and we write $J_s(w)$ instead of $J_s(w,h)$, etc.} For currents $J<J_c$ a stripe is magnetized uniformly within the stripe plane $x0y$ (perpendicularly to the current direction) and the magnetization direction has an angle $\phi=\phi(J)$ with the stripe direction ($x$-axis), see inset (b) in Fig.~\ref{fig:diagram}. The details of the dependence $\phi(J)$ will be discussed later. The appearance of the described inclined uniform in-plane state under the action of the current was recently predicted for the one-dimensional wires\cite{Kravchuk2013}.

For a certain thickness the saturation current $J_s(w)$ is a nonmonotonic function of the stripe width. The dependence $J_s(w)$ tends to zero in the limit case of very narrow stripes, $w\to0$. With the width increasing $J_s(w)$ rapidly reaches its maximum value and then, with the width further increasing, it demonstrates decaying oscillations which asymptotically approach the limit value $J_s^{\text{film}}$ as $w\to\infty$, where $J_s^{\text{film}}$ is the saturation current for an infinite film of the given thickness, see Fig.~\ref{fig:diagram}. The dependence $J_s^\mathrm{film}(h)$ was recently described both numerically and analytically, see Fig.~2 of Ref.~\onlinecite{Gaididei12a}.
	
The behavior of the current $J_c$ for narrow stripes is similar to the described behavior of the saturation current $J_s$ except that $J_c<J_s$. However, after passing the maximum the function $J_c(w)$ monotonically decays to zero.

There is an intermediate magnetization structure which appears for the range of the applied currents $J_c<J<J_s$ and drastically depends on the stripe width $w$. In this regard, one can distinguish two specific values of the width, namely $w_1\approx h$ and $w_2\approx2h$. For widths $w<w_1$ a stripe keeps the uniform magnetization for any value of the applied current. The transition from the uniform in-plane state to the saturation occurs by the appearance of an out-of-plane component $m_z$, see inset (c) in Fig.~\ref{fig:diagram}. This process is continuous for very narrow stripes ($w/h<\xi_c\approx 0.457$) and discontinuous for wider stripes $w/h>\xi_c$, see the analytical discussion below. The main feature of the range $w_1<w<w_2$ is that $J_c=J_s$, this means that the saturation is a \emph{discontinuous} process which occurs as a sharp jump from the uniform inclined state with $m_z=0$ to the saturated state with $m_z=1$. For the case $w>w_2$ the saturation occurs as a \emph{multilevel} process in which different intermediate periodic structures are possible. It is important to note that for small neighborhoods of the widths $w_1$ and $w_2$ (or in other words, for the cases $J_s\gtrapprox J_c$) the stable periodic structures appear in the intermediate regime $J_c<J<J_s$. For $w\approx w_1$ we find a one-dimensional domain structure, see Fig.~\ref{fig:diagram}(d), which coincides with the previously described domain structure arising in thin nanowires in the pre-saturated regime.\cite{Kravchuk2013} For $w\approx w_2$ we find another periodic structure, which we call the antivortex diamond state, see Fig.~\ref{fig:diagram}(e). This structure consists of a chain of antivortices aligned along the stripe central line, and two chains of edge solitons are aligned along the stripe edges. An edge soliton is half a vortex, or boojum.\cite{Volovik03}

Let us now consider the behavior of the relatively wide stripes, $w>w_2$. We start from the discussion of the pre-saturated regimes, where $J\lessapprox J_s$. In this case one observes two completely different stable states: namely the state of longitudinal domain walls (area 6 in Fig.~\ref{fig:diagram}) and the state of transverse domain walls (area 7 in Fig.~\ref{fig:diagram}). The first state appears just below bulges of the dependence $J_s(w)$ and represents a set of parallel Bloch walls aligned along the stripe; and the number of the domain walls increases with increasing width. The second state appears just below valleys of the dependence $J_s(w)$, it represents a periodic Bloch domain walls structure aligned perpendicularly to the stripe (along the $y$-axis). With the current decreasing the other periodic magnetization structures appears whose type is determined by the width: the cross-tie appears for the most narrow stripes ($w\gtrsim w_2$), see Fig.~\ref{fig:diagram}(f); the vortex diamond state appears for somewhat wider stripes, see Fig.~\ref{fig:diagram}(g); and the quasicrystal state appears for even wider samples, see Fig.~\ref{fig:diagram}(h). The cross-tie state is well described in the literature.\cite{Hubert98} The vortex diamond state coincides with the described antivortex one up to interchanging of positions of vortices and antivortices, this state was already observed in rectangular\cite{Hertel05,Hankemeier2009,Xie11} and elliptical\cite{Lai07} nanopaticles and also in mesoscopic islands of complicated form.\cite{Hertel05} The quasicrystal state is very similar to the vortex-antivortex lattice recently predicted for wide films\cite{Volkov11,Gaididei12a}. The periodic magnetization structures do not survive in wide stripes with further current decreasing, a dynamic regime of chaotic vortex-antivortex motion appears instead, see region 11 in Fig.~\ref{fig:diagram}.

It should be noted that occasionally defects can appear in the periodic structures. The defects are of vacancy or interstitial type with respect to vortices or antivortices.
	
The following analytical approach enables us to estimate the upper limit of the current $J_c$ and also the saturation current $J_s$ for the case of narrow stripes. For this purpose, let us consider a uniformly magnetized stripe under the action of the spin current. In this case the energy of the system consists only of the magnetostatic contribution and it can be expressed in the form\cite{Hubert98}
\begin{equation} \label{E_ms}
	\mathcal{E}=\mathcal{E}_{\mathrm{ms}} = \dfrac{V}{2} \left( N_x m^2_x + N_y m^2_y + N_z m^2_z \right),
\end{equation}
where $V$ is the sample volume and $N_x,~N_y,~N_z$ are demagnetizing factors which are determined by the stripe sizes, for details see Ref.~\onlinecite{Aharoni98}. For a long stripe $N_x\ll1$ and one can neglect it, however we save the factor $N_x$ to make our consideration more general.

Looking for the stationary uniform solutions, we substitute the expression~\eqref{E_ms} into the Eq.~\eqref{eq:LLS}, which results in the following set of equations
	\begin{subequations}\label{eq:system_1}
		\begin{align}
			\label{eq:mx}&\left( N_z - N_y\right) m_y m_z - j \varepsilon(m_z) m_x m_z = 0,\\
			\label{eq:my}&\left( N_x - N_z\right) m_x m_z - j \varepsilon(m_z) m_y m_z = 0,\\
			\label{eq:mz}&\left( N_y - N_x\right) m_x m_y - j \varepsilon(m_z) \left( m^2_z -1 \right) = 0.
		\end{align}
	\end{subequations}

It is easy to see that the set \eqref{eq:system_1} has an in-plane solution $m_z=0$, which turns the first two equations \eqref{eq:mx} and \eqref{eq:my} to identity and the last equation \eqref{eq:mz} enables one to obtain the orientation of the in-plane state:
\begin{equation} \label{eq:Sin_2phi}
	\sin(2\phi)=-J/\tilde J_c,
\end{equation}
where $\phi=\arctan(m_y/m_x)$ is an angle which the magnetization has with the $x$-axis, and
\begin{equation} \label{eq:J_c}
	\tilde J_c=J_0(1+\Lambda^{-2})(N_y-N_x)/(2P),
\end{equation}
is the highest current at which the uniform in-plane state exists. The critical current $\tilde J_c$ depends on the stripe sizes through the demagnetizing factors $N_y$ and $N_x$. The corresponding dependence $\tilde J_c(w)$ is shown in Fig.~\ref{fig:diagram} by a thin dotted line. As one can see $J_c=\tilde J_c$ for narrow stripes $\xi_ch\lesssim w\le h$, while for other values of the width $J_c<\tilde J_c$. This means that for a wide range of $w$ the instability of the uniform in-plane state occurs for currents lower than the value $\tilde J_c$, which can be only considered as an upper limit of the uniform in-plane state existence. However, the stability analysis is beyond the scope of the paper and will be performed in a further work.

The set of equations \eqref{eq:system_1} has also a solution with an out-of-plane component $m_z\ne0$:
\begin{subequations}\label{eq:sol-mz-ne0}
  \begin{align}
    m_z&=\frac{1}{1-\Lambda^{-2}}\left[2\frac{J}{\tilde J_s}-1-\Lambda^{-2}\right],\\
    \tan\phi&=\frac{m_y}{m_x} =\sqrt{(N_z-N_x)/(N_y-N_z)},
  \end{align}
\end{subequations}
where
\begin{equation}\label{eq:Js-tild}
  \tilde J_s=\frac{2J_0}{P}\sqrt{\left( N_y-N_z\right)\left( N_z-N_x\right)}
\end{equation}
is the current of transition to the saturated state $m_z=1$. The dependence $\tilde J_s$ is shown in Fig.~\ref{fig:diagram} by a black solid line and it demonstrates a very good agreement with the simulations data for $w<h$. The other limit condition $m_z\rightarrow0$ determines the current $\tilde J_c'=\tilde J_s(1+\Lambda^{-2})/2$, which in part separates the in-plane and out-of-plane uniform states, see dashed line in Fig.~\ref{fig:diagram}. Equality of the currents $\tilde J_c'=\tilde J_c$ is equivalent to the condition $N_z=1/3$ and it determines a critical aspect ratio $w/h=\xi_c\approx 0.457$, which separates continuous and discontinuous regimes of the saturation.

It is interesting to note that the solution \eqref{eq:sol-mz-ne0} requires the condition $N_y>N_z$ for the case of long stripes ($N_x\ll1$), i.e. the out-of-plane uniform state is possible only for wires with $w<h$.

In conclusion, we describe all possible types of the stripe magnetization behavior under the transverse current influence. The micromagnetic simulations were performed for all ranges of the currents, up to the saturation value. Considering a wide range of the stripe sizes we make a link between behavior of narrow one-dimensional and wide quasi two-dimensional systems. The saturation process is accompanied by appearance of stable periodic magnetization structures, the type of which changes with the stripe width: one-dimensional domain structure for narrow stripes, quasi two-dimensional vortex-antivortex lattice for wide stripes and cross-tie or diamond states for the intermediate range of the widths.

It is important to note than in contrast to wide samples, in stripes the vortex-antivortex lattice appears also for very small thicknesses, which means that a low value of the applied spin-current is needed. For example we found a  vortex-antivortex lattice in a stripe with thickness $h=5$ nm and the applied current was $J\approx4\times10^{12}~A/m^2$. This is much smaller than for a case of wide samples.\cite{Volkov11} So the spin-valves in form of long stripes are promising systems for an experimental observation of the vortex-antivortex lattices.

O.M.V. and V.P.K. thank the University of Bayreuth, where part of this work was performed, for kind hospitality. V.P.K acknowledges the support from DAAD (Codenumber A/13/03116) and Program of Fundamental Research of the Department of Physics and Astronomy of the NAS of Ukraine (project No. 0112U000056), O.M.V. acknowledges the support from the BAYHOST project.


\begin{thebibliography}{28}%
\makeatletter
\providecommand \@ifxundefined [1]{%
 \@ifx{#1\undefined}
}%
\providecommand \@ifnum [1]{%
 \ifnum #1\expandafter \@firstoftwo
 \else \expandafter \@secondoftwo
 \fi
}%
\providecommand \@ifx [1]{%
 \ifx #1\expandafter \@firstoftwo
 \else \expandafter \@secondoftwo
 \fi
}%
\providecommand \natexlab [1]{#1}%
\providecommand \enquote  [1]{``#1''}%
\providecommand \bibnamefont  [1]{#1}%
\providecommand \bibfnamefont [1]{#1}%
\providecommand \citenamefont [1]{#1}%
\providecommand \href@noop [0]{\@secondoftwo}%
\providecommand \href [0]{\begingroup \@sanitize@url \@href}%
\providecommand \@href[1]{\@@startlink{#1}\@@href}%
\providecommand \@@href[1]{\endgroup#1\@@endlink}%
\providecommand \@sanitize@url [0]{\catcode `\\12\catcode `\$12\catcode
  `\&12\catcode `\#12\catcode `\^12\catcode `\_12\catcode `\%12\relax}%
\providecommand \@@startlink[1]{}%
\providecommand \@@endlink[0]{}%
\providecommand \url  [0]{\begingroup\@sanitize@url \@url }%
\providecommand \@url [1]{\endgroup\@href {#1}{\urlprefix }}%
\providecommand \urlprefix  [0]{URL }%
\providecommand \Eprint [0]{\href }%
\providecommand \doibase [0]{http://dx.doi.org/}%
\providecommand \selectlanguage [0]{\@gobble}%
\providecommand \bibinfo  [0]{\@secondoftwo}%
\providecommand \bibfield  [0]{\@secondoftwo}%
\providecommand \translation [1]{[#1]}%
\providecommand \BibitemOpen [0]{}%
\providecommand \bibitemStop [0]{}%
\providecommand \bibitemNoStop [0]{.\EOS\space}%
\providecommand \EOS [0]{\spacefactor3000\relax}%
\providecommand \BibitemShut  [1]{\csname bibitem#1\endcsname}%
\let\auto@bib@innerbib\@empty
\bibitem [{\citenamefont {Lindner}(2010)}]{Lindner10}%
  \BibitemOpen
  \bibfield  {author} {\bibinfo {author} {\bibfnamefont {J.}~\bibnamefont
  {Lindner}},\ }\href
  {http://www.scopus.com/inward/record.url?eid=2-s2.0-77949267128&partnerID=40%
&md5=5bcc566c588a727451af4431fbd6a8bd} {\bibfield  {journal} {\bibinfo
  {journal} {Superlattices and Microstructures}\ }\textbf {\bibinfo {volume}
  {47}},\ \bibinfo {pages} {497} (\bibinfo {year} {2010})},\ \bibinfo {note}
  {cited By (since 1996) 1}\BibitemShut {NoStop}%
\bibitem [{\citenamefont {Tatara}, \citenamefont {Kohno},\ and\ \citenamefont
  {Shibata}(2008)}]{Tatara08}%
  \BibitemOpen
  \bibfield  {author} {\bibinfo {author} {\bibfnamefont {G.}~\bibnamefont
  {Tatara}}, \bibinfo {author} {\bibfnamefont {H.}~\bibnamefont {Kohno}}, \
  and\ \bibinfo {author} {\bibfnamefont {J.}~\bibnamefont {Shibata}},\ }\href
  {\doibase 10.1016/j.physrep.2008.07.003} {\bibfield  {journal} {\bibinfo
  {journal} {Physics Reports}\ }\textbf {\bibinfo {volume} {468}},\ \bibinfo
  {pages} {213} (\bibinfo {year} {2008})}\BibitemShut {NoStop}%
\bibitem [{\citenamefont {Marrows}(2005)}]{Marrows05}%
  \BibitemOpen
  \bibfield  {author} {\bibinfo {author} {\bibfnamefont {C.~H.}\ \bibnamefont
  {Marrows}},\ }\href {\doibase 10.1080/00018730500442209} {\bibfield
  {journal} {\bibinfo  {journal} {Advances in Physics}\ }\textbf {\bibinfo
  {volume} {54}},\ \bibinfo {pages} {585} (\bibinfo {year} {2005})}\BibitemShut
  {NoStop}%
\bibitem [{\citenamefont {Kl\"{a}ui}(2008)}]{Klaui08}%
  \BibitemOpen
  \bibfield  {author} {\bibinfo {author} {\bibfnamefont {M.}~\bibnamefont
  {Kl\"{a}ui}},\ }\href {http://stacks.iop.org/0953-8984/20/i=31/a=313001}
  {\bibfield  {journal} {\bibinfo  {journal} {Journal of Physics: Condensed
  Matter}\ }\textbf {\bibinfo {volume} {20}},\ \bibinfo {pages} {313001}
  (\bibinfo {year} {2008})}\BibitemShut {NoStop}%
\bibitem [{\citenamefont {Khvalkovskiy}\ \emph {et~al.}(2009)\citenamefont
  {Khvalkovskiy}, \citenamefont {Zvezdin}, \citenamefont {Gorbunov},
  \citenamefont {Cros}, \citenamefont {Grollier}, \citenamefont {Fert},\ and\
  \citenamefont {Zvezdin}}]{Khvalkovskiy09}%
  \BibitemOpen
  \bibfield  {author} {\bibinfo {author} {\bibfnamefont {A.~V.}\ \bibnamefont
  {Khvalkovskiy}}, \bibinfo {author} {\bibfnamefont {K.~A.}\ \bibnamefont
  {Zvezdin}}, \bibinfo {author} {\bibfnamefont {Y.~V.}\ \bibnamefont
  {Gorbunov}}, \bibinfo {author} {\bibfnamefont {V.}~\bibnamefont {Cros}},
  \bibinfo {author} {\bibfnamefont {J.}~\bibnamefont {Grollier}}, \bibinfo
  {author} {\bibfnamefont {A.}~\bibnamefont {Fert}}, \ and\ \bibinfo {author}
  {\bibfnamefont {A.~K.}\ \bibnamefont {Zvezdin}},\ }\href {\doibase
  10.1103/PhysRevLett.102.067206} {\bibfield  {journal} {\bibinfo  {journal}
  {Phys. Rev. Lett.}\ }\textbf {\bibinfo {volume} {102}},\ \bibinfo {eid}
  {067206} (\bibinfo {year} {2009})}\BibitemShut {NoStop}%
\bibitem [{\citenamefont {Boone}\ \emph {et~al.}(2010)\citenamefont {Boone},
  \citenamefont {Katine}, \citenamefont {Carey}, \citenamefont {Childress},
  \citenamefont {Cheng},\ and\ \citenamefont {Krivorotov}}]{Boone10a}%
  \BibitemOpen
  \bibfield  {author} {\bibinfo {author} {\bibfnamefont {C.~T.}\ \bibnamefont
  {Boone}}, \bibinfo {author} {\bibfnamefont {J.~A.}\ \bibnamefont {Katine}},
  \bibinfo {author} {\bibfnamefont {M.}~\bibnamefont {Carey}}, \bibinfo
  {author} {\bibfnamefont {J.~R.}\ \bibnamefont {Childress}}, \bibinfo {author}
  {\bibfnamefont {X.}~\bibnamefont {Cheng}}, \ and\ \bibinfo {author}
  {\bibfnamefont {I.~N.}\ \bibnamefont {Krivorotov}},\ }\href {\doibase
  10.1103/PhysRevLett.104.097203} {\bibfield  {journal} {\bibinfo  {journal}
  {Phys. Rev. Lett.}\ }\textbf {\bibinfo {volume} {104}},\ \bibinfo {pages}
  {097203} (\bibinfo {year} {2010})}\BibitemShut {NoStop}%
\bibitem [{\citenamefont {Uhl\'\i\ifmmode~\check{r}\else \v{r}\fi{}}\ \emph
  {et~al.}(2010)\citenamefont {Uhl\'\i\ifmmode~\check{r}\else \v{r}\fi{}},
  \citenamefont {Pizzini}, \citenamefont {Rougemaille}, \citenamefont
  {Novotn\'y}, \citenamefont {Cros}, \citenamefont {Jim\'enez}, \citenamefont
  {Faini}, \citenamefont {Heyne}, \citenamefont {Sirotti}, \citenamefont
  {Tieg}, \citenamefont {Bendounan}, \citenamefont {Maccherozzi}, \citenamefont
  {Belkhou}, \citenamefont {Grollier}, \citenamefont {Anane},\ and\
  \citenamefont {Vogel}}]{Uhlir2010}%
  \BibitemOpen
  \bibfield  {author} {\bibinfo {author} {\bibfnamefont {V.}~\bibnamefont
  {Uhl\'\i\ifmmode~\check{r}\else \v{r}\fi{}}}, \bibinfo {author}
  {\bibfnamefont {S.}~\bibnamefont {Pizzini}}, \bibinfo {author} {\bibfnamefont
  {N.}~\bibnamefont {Rougemaille}}, \bibinfo {author} {\bibfnamefont
  {J.}~\bibnamefont {Novotn\'y}}, \bibinfo {author} {\bibfnamefont
  {V.}~\bibnamefont {Cros}}, \bibinfo {author} {\bibfnamefont {E.}~\bibnamefont
  {Jim\'enez}}, \bibinfo {author} {\bibfnamefont {G.}~\bibnamefont {Faini}},
  \bibinfo {author} {\bibfnamefont {L.}~\bibnamefont {Heyne}}, \bibinfo
  {author} {\bibfnamefont {F.}~\bibnamefont {Sirotti}}, \bibinfo {author}
  {\bibfnamefont {C.}~\bibnamefont {Tieg}}, \bibinfo {author} {\bibfnamefont
  {A.}~\bibnamefont {Bendounan}}, \bibinfo {author} {\bibfnamefont
  {F.}~\bibnamefont {Maccherozzi}}, \bibinfo {author} {\bibfnamefont
  {R.}~\bibnamefont {Belkhou}}, \bibinfo {author} {\bibfnamefont
  {J.}~\bibnamefont {Grollier}}, \bibinfo {author} {\bibfnamefont
  {A.}~\bibnamefont {Anane}}, \ and\ \bibinfo {author} {\bibfnamefont
  {J.}~\bibnamefont {Vogel}},\ }\href {\doibase 10.1103/PhysRevB.81.224418}
  {\bibfield  {journal} {\bibinfo  {journal} {Phys. Rev. B}\ }\textbf {\bibinfo
  {volume} {81}},\ \bibinfo {pages} {224418} (\bibinfo {year}
  {2010})}\BibitemShut {NoStop}%
\bibitem [{\citenamefont {Chanthbouala}\ \emph {et~al.}(2011)\citenamefont
  {Chanthbouala}, \citenamefont {Matsumoto}, \citenamefont {Grollier},
  \citenamefont {Cros}, \citenamefont {Anane}, \citenamefont {Fert},
  \citenamefont {Khvalkovskiy}, \citenamefont {Zvezdin}, \citenamefont
  {Nishimura}, \citenamefont {Nagamine}, \citenamefont {Maehara}, \citenamefont
  {Tsunekawa}, \citenamefont {Fukushima},\ and\ \citenamefont
  {S.}}]{Chanthbouala11}%
  \BibitemOpen
  \bibfield  {author} {\bibinfo {author} {\bibfnamefont {A.}~\bibnamefont
  {Chanthbouala}}, \bibinfo {author} {\bibfnamefont {R.}~\bibnamefont
  {Matsumoto}}, \bibinfo {author} {\bibfnamefont {J.}~\bibnamefont {Grollier}},
  \bibinfo {author} {\bibfnamefont {V.}~\bibnamefont {Cros}}, \bibinfo {author}
  {\bibfnamefont {A.}~\bibnamefont {Anane}}, \bibinfo {author} {\bibfnamefont
  {A.}~\bibnamefont {Fert}}, \bibinfo {author} {\bibfnamefont {A.~V.}\
  \bibnamefont {Khvalkovskiy}}, \bibinfo {author} {\bibfnamefont {K.~A.}\
  \bibnamefont {Zvezdin}}, \bibinfo {author} {\bibfnamefont {K.}~\bibnamefont
  {Nishimura}}, \bibinfo {author} {\bibfnamefont {Y.}~\bibnamefont {Nagamine}},
  \bibinfo {author} {\bibfnamefont {H.}~\bibnamefont {Maehara}}, \bibinfo
  {author} {\bibfnamefont {K.}~\bibnamefont {Tsunekawa}}, \bibinfo {author}
  {\bibfnamefont {A.}~\bibnamefont {Fukushima}}, \ and\ \bibinfo {author}
  {\bibfnamefont {Y.}~\bibnamefont {S.}},\ }\href@noop {} {\bibfield  {journal}
  {\bibinfo  {journal} {Nature Physics}\ }\textbf {\bibinfo {volume} {7}},\
  \bibinfo {pages} {626} (\bibinfo {year} {2011})}\BibitemShut {NoStop}%
\bibitem [{\citenamefont {Metaxas}\ \emph {et~al.}(2013)\citenamefont
  {Metaxas}, \citenamefont {Sampaio}, \citenamefont {Chanthbouala},
  \citenamefont {Matsumoto}, \citenamefont {Anane}, \citenamefont {Fert},
  \citenamefont {Zvezdin}, \citenamefont {Yakushiji}, \citenamefont {Kubota},
  \citenamefont {Fukushima}, \citenamefont {Yuasa}, \citenamefont {Nishimura},
  \citenamefont {Nagamine}, \citenamefont {Maehara}, \citenamefont {Tsunekawa},
  \citenamefont {Cros},\ and\ \citenamefont {Grollier}}]{Metaxas2013}%
  \BibitemOpen
  \bibfield  {author} {\bibinfo {author} {\bibfnamefont {P.~J.}\ \bibnamefont
  {Metaxas}}, \bibinfo {author} {\bibfnamefont {J.}~\bibnamefont {Sampaio}},
  \bibinfo {author} {\bibfnamefont {A.}~\bibnamefont {Chanthbouala}}, \bibinfo
  {author} {\bibfnamefont {R.}~\bibnamefont {Matsumoto}}, \bibinfo {author}
  {\bibfnamefont {A.}~\bibnamefont {Anane}}, \bibinfo {author} {\bibfnamefont
  {A.}~\bibnamefont {Fert}}, \bibinfo {author} {\bibfnamefont {K.~A.}\
  \bibnamefont {Zvezdin}}, \bibinfo {author} {\bibfnamefont {K.}~\bibnamefont
  {Yakushiji}}, \bibinfo {author} {\bibfnamefont {H.}~\bibnamefont {Kubota}},
  \bibinfo {author} {\bibfnamefont {A.}~\bibnamefont {Fukushima}}, \bibinfo
  {author} {\bibfnamefont {S.}~\bibnamefont {Yuasa}}, \bibinfo {author}
  {\bibfnamefont {K.}~\bibnamefont {Nishimura}}, \bibinfo {author}
  {\bibfnamefont {Y.}~\bibnamefont {Nagamine}}, \bibinfo {author}
  {\bibfnamefont {H.}~\bibnamefont {Maehara}}, \bibinfo {author} {\bibfnamefont
  {K.}~\bibnamefont {Tsunekawa}}, \bibinfo {author} {\bibfnamefont
  {V.}~\bibnamefont {Cros}}, \ and\ \bibinfo {author} {\bibfnamefont
  {J.}~\bibnamefont {Grollier}},\ }\href {http://dx.doi.org/10.1038/srep01829}
  {\bibfield  {journal} {\bibinfo  {journal} {Sci. Rep.}\ }\textbf {\bibinfo
  {volume} {3}},\ \bibinfo {pages} {1829} (\bibinfo {year} {2013})}\BibitemShut
  {NoStop}%
\bibitem [{\citenamefont {Volkov}\ \emph {et~al.}(2011)\citenamefont {Volkov},
  \citenamefont {Kravchuk}, \citenamefont {Sheka},\ and\ \citenamefont
  {Gaididei}}]{Volkov11}%
  \BibitemOpen
  \bibfield  {author} {\bibinfo {author} {\bibfnamefont {O.~M.}\ \bibnamefont
  {Volkov}}, \bibinfo {author} {\bibfnamefont {V.~P.}\ \bibnamefont
  {Kravchuk}}, \bibinfo {author} {\bibfnamefont {D.~D.}\ \bibnamefont {Sheka}},
  \ and\ \bibinfo {author} {\bibfnamefont {Y.}~\bibnamefont {Gaididei}},\
  }\href {\doibase 10.1103/PhysRevB.84.052404} {\bibfield  {journal} {\bibinfo
  {journal} {Phys. Rev. B}\ }\textbf {\bibinfo {volume} {84}},\ \bibinfo
  {pages} {052404} (\bibinfo {year} {2011})}\BibitemShut {NoStop}%
\bibitem [{\citenamefont {Gaididei}\ \emph {et~al.}(2012)\citenamefont
  {Gaididei}, \citenamefont {Volkov}, \citenamefont {Kravchuk},\ and\
  \citenamefont {Sheka}}]{Gaididei12a}%
  \BibitemOpen
  \bibfield  {author} {\bibinfo {author} {\bibfnamefont {Y.}~\bibnamefont
  {Gaididei}}, \bibinfo {author} {\bibfnamefont {O.~M.}\ \bibnamefont
  {Volkov}}, \bibinfo {author} {\bibfnamefont {V.~P.}\ \bibnamefont
  {Kravchuk}}, \ and\ \bibinfo {author} {\bibfnamefont {D.~D.}\ \bibnamefont
  {Sheka}},\ }\href {\doibase 10.1103/PhysRevB.86.144401} {\bibfield  {journal}
  {\bibinfo  {journal} {Phys. Rev. B}\ }\textbf {\bibinfo {volume} {86}},\
  \bibinfo {pages} {144401} (\bibinfo {year} {2012})}\BibitemShut {NoStop}%
\bibitem [{\citenamefont {Kravchuk}\ \emph {et~al.}(2013)\citenamefont
  {Kravchuk}, \citenamefont {Volkov}, \citenamefont {Sheka},\ and\
  \citenamefont {Gaididei}}]{Kravchuk2013}%
  \BibitemOpen
  \bibfield  {author} {\bibinfo {author} {\bibfnamefont {V.~P.}\ \bibnamefont
  {Kravchuk}}, \bibinfo {author} {\bibfnamefont {O.~M.}\ \bibnamefont
  {Volkov}}, \bibinfo {author} {\bibfnamefont {D.~D.}\ \bibnamefont {Sheka}}, \
  and\ \bibinfo {author} {\bibfnamefont {Y.}~\bibnamefont {Gaididei}},\ }\href
  {\doibase 10.1103/PhysRevB.87.224402} {\bibfield  {journal} {\bibinfo
  {journal} {Phys. Rev. B}\ }\textbf {\bibinfo {volume} {87}},\ \bibinfo
  {pages} {224402} (\bibinfo {year} {2013})}\BibitemShut {NoStop}%
\bibitem [{\citenamefont {Slonczewski}(1996)}]{Slonczewski96}%
  \BibitemOpen
  \bibfield  {author} {\bibinfo {author} {\bibfnamefont {J.~C.}\ \bibnamefont
  {Slonczewski}},\ }\href {http://dx.doi.org/10.1016/0304-8853(96)00062-5}
  {\bibfield  {journal} {\bibinfo  {journal} {J.~Magn. Magn. Mater.}\ }\textbf
  {\bibinfo {volume} {159}},\ \bibinfo {pages} {L1} (\bibinfo {year}
  {1996})}\BibitemShut {NoStop}%
\bibitem [{\citenamefont {Berger}(1996)}]{Berger96}%
  \BibitemOpen
  \bibfield  {author} {\bibinfo {author} {\bibfnamefont {L.}~\bibnamefont
  {Berger}},\ }\href {http://link.aps.org/abstract/PRB/v54/p9353} {\bibfield
  {journal} {\bibinfo  {journal} {Phys. Rev. B}\ }\textbf {\bibinfo {volume}
  {54}},\ \bibinfo {pages} {9353} (\bibinfo {year} {1996})}\BibitemShut
  {NoStop}%
\bibitem [{\citenamefont {Slonczewski}(2002)}]{Slonczewski02}%
  \BibitemOpen
  \bibfield  {author} {\bibinfo {author} {\bibfnamefont {J.~C.}\ \bibnamefont
  {Slonczewski}},\ }\href
  {http://www.sciencedirect.com/science/article/B6TJJ-45RFP8K-5/2/74b7bbf5e7dc%
831281ad65874ccb88b9} {\bibfield  {journal} {\bibinfo  {journal} {J.~Magn.
  Magn. Mater.}\ }\textbf {\bibinfo {volume} {247}},\ \bibinfo {pages} {324}
  (\bibinfo {year} {2002})}\BibitemShut {NoStop}%
\bibitem [{\citenamefont {Volkov}\ and\ \citenamefont
  {Kravchuk}(2013)}]{Volkov2013}%
  \BibitemOpen
  \bibfield  {author} {\bibinfo {author} {\bibfnamefont {A.}~\bibnamefont
  {Volkov}}\ and\ \bibinfo {author} {\bibfnamefont {V.}~\bibnamefont
  {Kravchuk}},\ }\href@noop {} {\bibfield  {journal} {\bibinfo  {journal}
  {Ukrainian Journal of Physics}\ }\textbf {\bibinfo {volume} {58}},\ \bibinfo
  {pages} {667} (\bibinfo {year} {2013})}\BibitemShut {NoStop}%
\bibitem [{\citenamefont {Sluka}\ \emph {et~al.}(2011)\citenamefont {Sluka},
  \citenamefont {K{\'a}kay}, \citenamefont {Deac}, \citenamefont {B{\"u}rgler},
  \citenamefont {Hertel},\ and\ \citenamefont {Schneider}}]{Sluka11}%
  \BibitemOpen
  \bibfield  {author} {\bibinfo {author} {\bibfnamefont {V.}~\bibnamefont
  {Sluka}}, \bibinfo {author} {\bibfnamefont {A.}~\bibnamefont {K{\'a}kay}},
  \bibinfo {author} {\bibfnamefont {A.~M.}\ \bibnamefont {Deac}}, \bibinfo
  {author} {\bibfnamefont {D.~E.}\ \bibnamefont {B{\"u}rgler}}, \bibinfo
  {author} {\bibfnamefont {R.}~\bibnamefont {Hertel}}, \ and\ \bibinfo {author}
  {\bibfnamefont {C.~M.}\ \bibnamefont {Schneider}},\ }\href
  {http://stacks.iop.org/0022-3727/44/i=38/a=384002} {\bibfield  {journal}
  {\bibinfo  {journal} {Journal of Physics D: Applied Physics}\ }\textbf
  {\bibinfo {volume} {44}},\ \bibinfo {pages} {384002} (\bibinfo {year}
  {2011})}\BibitemShut {NoStop}%
\bibitem [{Note1()}]{Note1}%
  \BibitemOpen
  \bibinfo {note} {We use the OOMMF code, version 1.2a5
  [http://math.nist.gov/oommf/] for material parameters of Permalloy ($\protect
  \mathrm {Ni}_{81}\protect \mathrm {Fe}_{19}$): saturation magnetization
  $M_s=8.6 \times 10^5$ A/m, exchange constant $A=13 \times 10^{-12}$ J/m, and
  anisotropy is neglected. Size of the mesh cell is $3 \times X \times h$ nm,
  where $X$ takes values in interval from 2 to 3 nm, depending on the stripe
  width. The width is changes with steps $\Delta w=0.5$~nm for narrow stripes
  (0.5$\le w\le 5$~nm) and $\Delta w=1$~nm for other samples (5$< w\le
  100$~nm). The current parameters are the following: polarization degree $P =
  0.4$, and $\Lambda = 2$.}\BibitemShut {Stop}%
\bibitem [{Note2()}]{Note2}%
  \BibitemOpen
  \bibinfo {note} {Density of the applied current is changed accordingly to the
  law: $J=t~\Delta J/\Delta t$, where $\Delta J=10^{11}~A/m^2$ and $\Delta
  t=1~ns$. As a criterion of the saturation we use the relation
  $M_z/M_s>0.9999$, where $M_z$ is the total magnetization along the
  $z$-axis.}\BibitemShut {Stop}%
\bibitem [{Note3()}]{Note3}%
  \BibitemOpen
  \bibinfo {note} {Since the diagram of states (see Fig.~\ref {fig:diagram}) is
  built for a certain thickness, we omit the dependence on $h$ and we write
  $J_s(w)$ instead of $J_s(w,h)$, etc.}\BibitemShut {Stop}%
\bibitem [{\citenamefont {Hertel}\ \emph {et~al.}(2007)\citenamefont {Hertel},
  \citenamefont {Gliga}, \citenamefont {F\"ahnle},\ and\ \citenamefont
  {Schneider}}]{Hertel07}%
  \BibitemOpen
  \bibfield  {author} {\bibinfo {author} {\bibfnamefont {R.}~\bibnamefont
  {Hertel}}, \bibinfo {author} {\bibfnamefont {S.}~\bibnamefont {Gliga}},
  \bibinfo {author} {\bibfnamefont {M.}~\bibnamefont {F\"ahnle}}, \ and\
  \bibinfo {author} {\bibfnamefont {C.~M.}\ \bibnamefont {Schneider}},\ }\href
  {http://link.aps.org/abstract/PRL/v98/e117201} {\bibfield  {journal}
  {\bibinfo  {journal} {Phys. Rev. Lett.}\ }\textbf {\bibinfo {volume} {98}},\
  \bibinfo {eid} {117201} (\bibinfo {year} {2007})}\BibitemShut {NoStop}%
\bibitem [{\citenamefont {Volovik}(2003)}]{Volovik03}%
  \BibitemOpen
  \bibfield  {author} {\bibinfo {author} {\bibfnamefont {G.}~\bibnamefont
  {Volovik}},\ }\href
  {http://www.zentralblatt-math.org/zmath/search/?an=01866310} {\emph {\bibinfo
  {title} {The universe in a {H}elium droplet}}}\ (\bibinfo  {publisher}
  {Oxford University Press},\ \bibinfo {address} {Oxford},\ \bibinfo {year}
  {2003})\BibitemShut {NoStop}%
\bibitem [{\citenamefont {Hubert}\ and\ \citenamefont {Sch{\"
  a}fer}(1998)}]{Hubert98}%
  \BibitemOpen
  \bibfield  {author} {\bibinfo {author} {\bibfnamefont {A.}~\bibnamefont
  {Hubert}}\ and\ \bibinfo {author} {\bibfnamefont {R.}~\bibnamefont {Sch{\"
  a}fer}},\ }\href@noop {} {\emph {\bibinfo {title} {Magnetic domains: the
  analysis of magnetic microstructures}}}\ (\bibinfo  {publisher}
  {Springer--Verlag},\ \bibinfo {address} {Berlin},\ \bibinfo {year}
  {1998})\BibitemShut {NoStop}%
\bibitem [{\citenamefont {Hertel}\ \emph {et~al.}(2005)\citenamefont {Hertel},
  \citenamefont {Fruchart}, \citenamefont {Cherifi}, \citenamefont {Jubert},
  \citenamefont {Heun}, \citenamefont {Locatelli},\ and\ \citenamefont
  {Kirschner}}]{Hertel05}%
  \BibitemOpen
  \bibfield  {author} {\bibinfo {author} {\bibfnamefont {R.}~\bibnamefont
  {Hertel}}, \bibinfo {author} {\bibfnamefont {O.}~\bibnamefont {Fruchart}},
  \bibinfo {author} {\bibfnamefont {S.}~\bibnamefont {Cherifi}}, \bibinfo
  {author} {\bibfnamefont {P.-O.}\ \bibnamefont {Jubert}}, \bibinfo {author}
  {\bibfnamefont {S.}~\bibnamefont {Heun}}, \bibinfo {author} {\bibfnamefont
  {A.}~\bibnamefont {Locatelli}}, \ and\ \bibinfo {author} {\bibfnamefont
  {J.}~\bibnamefont {Kirschner}},\ }\href {\doibase 10.1103/PhysRevB.72.214409}
  {\bibfield  {journal} {\bibinfo  {journal} {Phys. Rev. B}\ }\textbf {\bibinfo
  {volume} {72}},\ \bibinfo {eid} {214409} (\bibinfo {year}
  {2005})}\BibitemShut {NoStop}%
\bibitem [{\citenamefont {Hankemeier}\ \emph {et~al.}(2009)\citenamefont
  {Hankemeier}, \citenamefont {Fr\"omter}, \citenamefont {Mikuszeit},
  \citenamefont {Stickler}, \citenamefont {Stillrich}, \citenamefont
  {P\"utter}, \citenamefont {Vedmedenko},\ and\ \citenamefont
  {Oepen}}]{Hankemeier2009}%
  \BibitemOpen
  \bibfield  {author} {\bibinfo {author} {\bibfnamefont {S.}~\bibnamefont
  {Hankemeier}}, \bibinfo {author} {\bibfnamefont {R.}~\bibnamefont
  {Fr\"omter}}, \bibinfo {author} {\bibfnamefont {N.}~\bibnamefont
  {Mikuszeit}}, \bibinfo {author} {\bibfnamefont {D.}~\bibnamefont {Stickler}},
  \bibinfo {author} {\bibfnamefont {H.}~\bibnamefont {Stillrich}}, \bibinfo
  {author} {\bibfnamefont {S.}~\bibnamefont {P\"utter}}, \bibinfo {author}
  {\bibfnamefont {E.~Y.}\ \bibnamefont {Vedmedenko}}, \ and\ \bibinfo {author}
  {\bibfnamefont {H.~P.}\ \bibnamefont {Oepen}},\ }\href {\doibase
  10.1103/PhysRevLett.103.147204} {\bibfield  {journal} {\bibinfo  {journal}
  {Phys. Rev. Lett.}\ }\textbf {\bibinfo {volume} {103}},\ \bibinfo {pages}
  {147204} (\bibinfo {year} {2009})}\BibitemShut {NoStop}%
\bibitem [{\citenamefont {Xie}\ \emph {et~al.}(2011)\citenamefont {Xie},
  \citenamefont {Zhang}, \citenamefont {Lin}, \citenamefont {Zhang},\ and\
  \citenamefont {Sang}}]{Xie11}%
  \BibitemOpen
  \bibfield  {author} {\bibinfo {author} {\bibfnamefont {K.}~\bibnamefont
  {Xie}}, \bibinfo {author} {\bibfnamefont {X.}~\bibnamefont {Zhang}}, \bibinfo
  {author} {\bibfnamefont {W.}~\bibnamefont {Lin}}, \bibinfo {author}
  {\bibfnamefont {P.}~\bibnamefont {Zhang}}, \ and\ \bibinfo {author}
  {\bibfnamefont {H.}~\bibnamefont {Sang}},\ }\href {\doibase
  10.1103/PhysRevB.84.054460} {\bibfield  {journal} {\bibinfo  {journal} {Phys.
  Rev. B}\ }\textbf {\bibinfo {volume} {84}},\ \bibinfo {pages} {054460}
  (\bibinfo {year} {2011})}\BibitemShut {NoStop}%
\bibitem [{\citenamefont {Lai}\ \emph {et~al.}(2007)\citenamefont {Lai},
  \citenamefont {Wei}, \citenamefont {Wu}, \citenamefont {Shieh}, \citenamefont
  {Chang},\ and\ \citenamefont {Guo}}]{Lai07}%
  \BibitemOpen
  \bibfield  {author} {\bibinfo {author} {\bibfnamefont {M.-F.}\ \bibnamefont
  {Lai}}, \bibinfo {author} {\bibfnamefont {Z.-H.}\ \bibnamefont {Wei}},
  \bibinfo {author} {\bibfnamefont {J.~C.}\ \bibnamefont {Wu}}, \bibinfo
  {author} {\bibfnamefont {W.~Z.}\ \bibnamefont {Shieh}}, \bibinfo {author}
  {\bibfnamefont {C.~R.}\ \bibnamefont {Chang}}, \ and\ \bibinfo {author}
  {\bibfnamefont {J.}~\bibnamefont {Guo}},\ }\href {\doibase 10.1063/1.2714667}
  {\bibfield  {journal} {\bibinfo  {journal} {J.~Appl. Phys.}\ }\textbf
  {\bibinfo {volume} {101}},\ \bibinfo {eid} {09N111} (\bibinfo {year}
  {2007})}\BibitemShut {NoStop}%
\bibitem [{\citenamefont {Aharoni}(1998)}]{Aharoni98}%
  \BibitemOpen
  \bibfield  {author} {\bibinfo {author} {\bibfnamefont {A.}~\bibnamefont
  {Aharoni}},\ }\href {http://dx.doi.org/10.1063/1.367113} {\bibfield
  {journal} {\bibinfo  {journal} {Journal of Applied Physics}\ }\textbf
  {\bibinfo {volume} {83}},\ \bibinfo {pages} {3432} (\bibinfo {year}
  {1998})}\BibitemShut {NoStop}%
\end{thebibliography}

%

\end{document}